\documentclass[acmsmall]{acmart}

\usepackage{algorithm}
\usepackage{algpseudocode}

\renewcommand {\algorithmiccomment}{\hfill {//}}

\usepackage[mathscr]{eucal}
\usepackage{amsmath}  
\usepackage{witharrows}
\usepackage{makecell}
\usepackage{stfloats}
\usepackage{booktabs}
\usepackage{ulem} 
\usepackage{multicol}
\usepackage{hyperref}
\usepackage[hyphenbreaks]{breakurl}
\usepackage{multirow}
\usepackage{bbding}
\usepackage{graphicx}
\usepackage{subfigure}
\usepackage{float}
\usepackage{xcolor}
\newcommand{\add}[1]{{\color{black} #1}}

\begin{document}
\def\methodName#1{CompTuner{#1}}

\title{Compiler Auto-tuning through Multiple Phase Learning}

\author{Mingxuan Zhu}
\email{zhumingxuan@stu.pku.edu.cn}
\affiliation{%
  \institution{Key Laboratory of High Confidence Software Technologies (Peking University), Ministry
of Education}
  \streetaddress{No.5 Yiheyuan Road}
  \city{Beijing}
  \country{China}
  \postcode{100871}
}

\author{Dan Hao}
\affiliation{%
  \institution{Key Laboratory of High Confidence Software Technologies (Peking University), Ministry
of Education}
  \streetaddress{No.5 Yiheyuan Road}
  \city{Beijing}
  \country{China}
  \postcode{100871}
  }
\email{haodan@pku.edu.cn}

\author{Junjie Chen}
\affiliation{%
  \institution{College of Intelligence and Computing, Tianjin University}
  \streetaddress{No. 135 Yaguan Road}
  \postcode{300350}
  \city{Tianjin}
  \country{China}
}
\email{junjiechen@tju.edu.cn}

\renewcommand{\shortauthors}{Mingxuan Zhu et al.}

\begin{abstract}
Widely used compilers like GCC and LLVM usually have hundreds of optimizations controlled by optimization flags, which are enabled or disabled during compilation to improve runtime performance (e.g., small execution time) of the compiler program. 
Due to the large number of optimization flags and their combination, it is difficult for compiler users to manually tune compiler optimization flags. In the literature, a number of auto-tuning techniques have been proposed, which tune optimization flags for a compiled program by comparing its actual runtime performance with different optimization flag combination. 
Due to the huge search space and heavy actual runtime cost, these techniques suffer from the widely-recognized efficiency problem. 
To reduce the heavy runtime cost, in this paper we propose a lightweight learning approach which uses a small number of actual runtime performance data to predict the runtime performance of a compiled program with various optimization flag combination. 
Furthermore, to reduce the search space, we design a novel particle swarm algorithm which tunes compiler optimization flags with the prediction model. To evaluate the performance of the proposed approach \methodName, we conduct an extensive experimental study on two popular C compilers GCC and LLVM with two widely used benchmarks cBench and PolyBench. 
The experimental results show that \methodName~significantly outperforms the 
\add{five }compared techniques, including the state-of-art technique BOCA. 
\end{abstract}

\acmJournal{TOSEM}
\acmVolume{1}
\acmNumber{1}
\acmArticle{1}
\acmMonth{10}

\maketitle

\section{Introduction}
Compilers play an important role in software development, especially program execution. 
A widely used compiler like GCC or LLVM usually have hundreds of optimizations, e.g., \textbf{function inlining}, \textbf{useless code removal}, which are controlled by optimization flags. 
During compilation, these optimizations can be enabled or disabled, resulting in various compiled programs with different runtime performance~\cite{schneck1973survey,tiwari2009scalable,padua1986advanced}. 
For example, the compiler optimization flag \textbf{-floop-unroll-and-jam} can perform external loop unfolding and internal loop fusion to optimize loop statements in a program. That is, compiler optimization flags have a significant impact on the runtime performance of a compiled program. 

On the other hand, the same optimization does not always result in the same performance improvement when being applied to different programs. Although widely used compilers like GCC and LLVM have recommended optimization settings such as -O1, -O2, and -O3, such optimization settings do not guarantee to always improve runtime performance for any program. 

Therefore, it is necessary to choose \textbf{specific} compiler optimization settings for various programs, especially for the programs that need a long time to run and optimization flags have a significant impact on. 
\add{In particular, a compiler optimization setting (i.e., compiler tuning) consists of selecting a right set of compiler optimizations (also known as ``flag selection''~\cite{chen2021efficient, garciarena2016evolutionary} or ``phase selection''~\cite{DBLP:conf/vee/JantzK13}) and deciding the order in which these optimizations are applied (also known as ``phase ordering''~\cite{ashouri2017micomp, huang2019autophase}). In this paper, we focus on the former one, the flag selection problem.}
Considering the large number of optimizations and their exponential combinations, it is difficult for compiler users to understand the impact of each compiler flag on the program and manually choose compiler optimization settings (i.e., which can be also viewed as optimization sequences) for a program. 

To alleviate human efforts in setting compiler optimizations, In the literature a number of compiler auto-tuning techniques \add{focusing on \textbf{flag selection}} have been proposed, which are divided into two categories~\cite{ashouri2018survey,ashouri2016compiler}, supervised learning based and unsupervised learning based techniques. 
In particular, unsupervised learning based techniques~\add{\cite{perez2017automatic,hoste2008cole,sandran2012optimized,garciarena2016evolutionary,purini2013finding,ni2019evolutionary,almagor2004finding,monsifrot2002machine,cavazos2007rapidly}} were firstly proposed to address compiler auto-tuning. Typically, for a program under compiling, the unsupervised learning based techniques explore its searching space of optimization sequences through some searching strategies (e.g., hill climbing based algorithm~\cite{almagor2004finding} and genetic algorithm~\cite{hoste2008cole,sandran2012optimized,garciarena2016evolutionary,ni2019evolutionary}), and select optimization sequences based on the actual runtime performance of the program compiled with the corresponding optimization sequences. During the searching process, these techniques usually generate a large number of optimization sequences, whose corresponding runtime performance of the target program has to be collected via compilation and execution. That is, these techniques are very time-consuming.
To alleviate the cost issue, supervised learning based techniques have been proposed, which first build a model to predict runtime performance of an optimization sequence, and use some searching strategies to find a desired optimization sequence based on the predicted runtime performance instead of actual performance. In particular, these techniques~\cite{chen2021efficient,park2013predictive,ashouri2014bayesian,cavazos2006automatic,ashouri2018automatic,chen2012deconstructing,stephenson2003meta} build a prediction model with a large number of data (i.e., optimization sequences and the actual runtime performance of a program compiled with the corresponding optimization sequences), and thus they still suffer from the cost issue. Furthermore, due to the huge combination space of optimization flags, it is always challenging to find a desired optimization sequence and the existing compiler auto-tuning techniques (including both supervised and unsupervised learning based techniques) still suffer from the performance issue. For example, our experimental study shows that the state-of-art unsupervised learning based technique (i.e., GA-based technique~\cite{garciarena2016evolutionary}) achieves the desired optimization results on only 2/40 programs, and the state-of-art supervised learning technique BOCA~\cite{chen2021efficient}\footnote{ \add{BOCA is the latest technique on compiler auto-tuning published in ICSE 2021. It reduces the search space by identifying impactful optimization flags by the prediction model, and selects the optimization sequences with the highest \textbf{expected improvement} iteratively to tune the prediction model.}} still spends almost three times longer to achieve the optimized runtime performance of random optimization on program \textbf{telecom\_adpcm\_c}. 

To alleviate the time cost and performance issues of existing compiler auto-tuning techniques, in this paper, we propose a multiple-phase learning based compiler auto-tuning technique \textbf{\methodName}, which first builds a prediction model through multiple phases of learning and searches for a desired optimization sequence through another round of learning. In particular, \methodName~first builds a prediction model through multiple phases of learning, each of which uses only a very small number of carefully selected data (i.e., optimization sequences and the corresponding actual runtime performance). This lightweight learning process alleviates the cost issue of existing supervised learning based techniques without scarifying the accuracy of a prediction model. Then, \methodName~searches for an optimization sequence with good runtime performance through an improved particle swarm optimization algorithm~\add{\cite{kennedy1995particle,shi1999empirical,poli2007particle}}, which balances the local exploitation and global exploration considering performance diversity. \add{The proposed \methodName~is different from the latest work BOCA in both building prediction models and the searching process. In particular, BOCA builds a prediction model through selected sequences with the best performance, whereas CompTuner builds a prediction model by considering sequences with performance diversity. In the searching process, BOCA narrows the search space by identifying impactful flags while CompTuner searches with an improved particle swarm algorithm.
}

To evaluate the performance of our approach, we conducted an experiment on two widely-used compilers GCC 8.3.0 and LLVM 9.0.0 using 20 programs from two widely benchmarks cBench and PolyBench. 
From the experimental results, the optimization sequences generated by \methodName~are always better than the default -O3 optimization of GCC and LLVM. 
\add{Moreover, given a time limitation for compiler auto-tuning, the optimization sequences generated by \methodName~significantly outperform all the five representative compiler auto-tuning techniques. Moreover, our proposed \methodName~achieves the best optimization results on many cases (i.e., 8 programs on GCC and 8 programs on LLVM). Besides, we manually analyze some optimization flags selected by BOCA, \methodName, as well as the best configuration obtained through a searching process, and find that BOCA recommends different optimization flags for a program in different trials, whereas the optimization flags selected by \methodName~and the best configuration have large overlap.

}

Furthermore, from the ablation study, the prediction model built by \methodName~has high accuracy (whose lowest prediction error rate is only range 1.49\% from while using 60 training data), resulting in the good acceleration performance of \methodName. 
Moreover, the improved particle swarm optimization algorithm does improve the performance of \methodName~by selecting optimization sequence with diversity. Among the 6 experimental cases we chose, \methodName~achieves better speedups on 4 cases, while the variant technique\footnote{That is, a variant technique uses impactful flags to reduce the search space instead of particle swarm optimization.} only performs better on 2 cases. 

The contributions of this paper are summarized as below.
\begin{itemize}
  \item \textbf{A novel multiple-phase learning based compiler auto-tuning approach \methodName}, which builds a prediction model with small numbers of selected instances and recommends compiler optimization sequences through improved particle swarm optimization.
  \item \textbf{An extensive experiment} on GCC and LLVM using two compiler optimization benchmarks cBench and PolyBench, which demonstrates that \methodName~is effective and promising.
  \item \textbf{A reproducible package} available at:~
  \href{https://github.com/TOSEM202210/compTuner}{$https://github.com/TOSEM202210/compTuner$}
\end{itemize}

The remaining of this paper is organized as follows. Section~\ref{sec:approach} introduces the approach details. Section~\ref{sec:setup} presents the experimental setup and Section~\ref{sec:results} presents the experimental results and analysis. Section~\ref{sec:threats} presents the threats to validity. \add{Section~\ref{sec:discuss} presents the discussion from three aspects.} Section~\ref{sec:related} briefly reviews the related work. Finally, Section~\ref{sec:conclude} concludes the whole paper. 

\section{Technique}
\label{sec:approach}

\begin{figure*}[h]
  \centering
  \includegraphics[width=\linewidth]{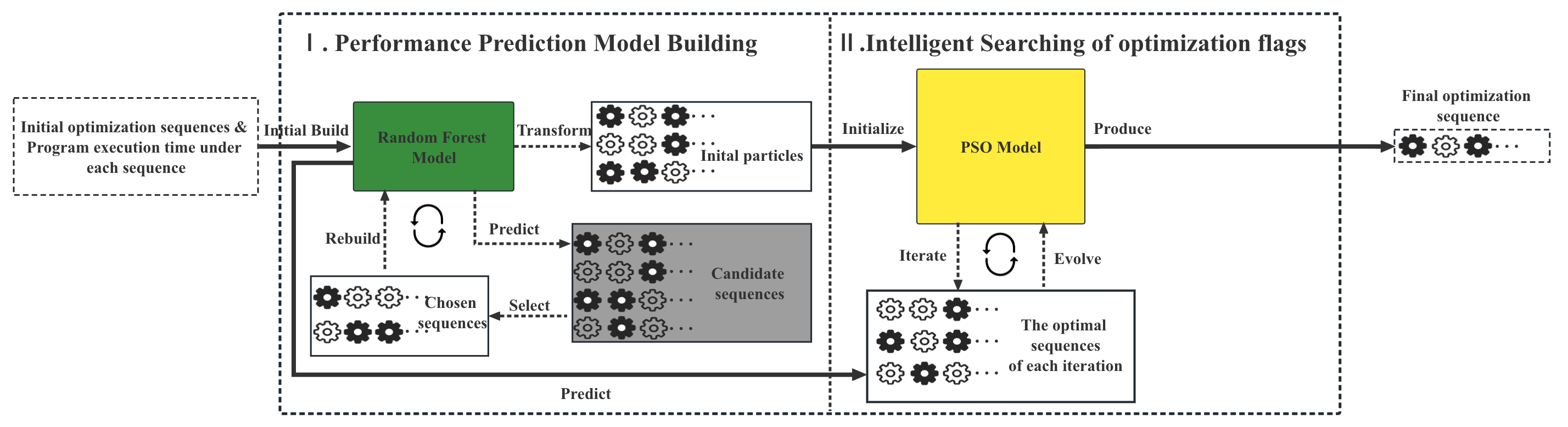}
  \caption{Workflow of \methodName.}
  \label{fig}
  \Description{Workflow of the approach.}
\end{figure*}

A compiler is to translate a program in one programming language into another programming language, often resulting in an executable program. To improve the compiled-code size or runtime performance, a traditional compiler like GCC or LLVM usually has hundreds of compiler optimization flags. For example, the GCC's  optimization flags \textbf{-floop-interchange}, \textbf{-floop-unroll-and-jam}, and \textbf{-fmove-loop-invariants} are to optimize the loop part of a program, and \textbf{-fcse-follow-jumps} and  \textbf{-fguess-branch-probability} are to optimize the branch part of a program. In particular, an optimization flag specifies a kind of program transformation for some program structure, whose value is usually 0/1\footnote{A few optimization flags have other values instead of 0 and 1. For example, \textbf{-falign-functions} can be set to 24 or 32, which represents the align position of a function. Following all previous work, we focus on optimization flags whose values are only 0 or 1 in this paper.} (i.e., 0 represents that the corresponding optimization flag is disabled and 1 represents that the corresponding optimization flag is enabled). 

For ease of presentation, we formalize the problem this paper targets as below. Given a compiler, its set of optimizations can be represented by $\mathcal{O} = \{o_1,o_2,...,o_n\}$, where $n$ represents the number of optimizations and the value of $o_i$ (where $1 \leq i \leq n$) represents whether the corresponding optimization flag is enabled or disabled (i.e., $o_i$ is set 1 or 0). A specific setting on these optimization flags can be represented by a 0/1 sequence, which is called an optimization sequence~\cite{chen2021efficient}. 
The size of the whole optimization space (i.e., including all optimization sequences) is $2^n$, which is very huge. For example, GCC 8.3.0 has 106 optimizations, and the size of the optimization space is $2^{106}$. On the other hand, enabling an optimization flag does not always improve the runtime behavior of a compiled program and the same optimization flag may have different affects on various compiled programs. Due to the huge optimization space and diverse performance of each optimization flag, manually compiler tuning is impossible, and thus in this paper we target at compiler auto-tuning, i.e., automatically identifying a specific optimization sequence for a compiled program with the goal of runtime performance improvement. 
In particular, this paper focuses on short execution time, which is the mostly-used compiler optimization goal~\cite{chen2021efficient}, and leave other compiler optimization goals like small executable code as our future work.

\subsection{Approach Overview}
\label{sec:overview}
The existing compiler auto-tuning techniques usually find a desired optimization sequence by comparing the \textit{actual} execution time of the compiled program with different optimization sequences (i.e., setting various values to the optimization flags in $O$), suffering the performance problem due to the huge search space and execution time. To address the performance problem, in this paper, we propose a multiple phase learning based approach \methodName, which alleviates the two performance issues (i.e., long execution time and huge search space) through two learning phases.

Figure \ref{fig} shows the overview of \methodName, which consists of prediction model building and intelligent searching.
\begin{itemize}
    \item \textit{Prediction model building} is to learn a prediction model on the runtime performance of a compiled program with various optimization sequences by using only a small number of training instances. In particular, we propose a lightweight learning approach, which builds a prediction model through initial learning on randomly generated optimization sequences and enhances the prediction accuracy through several phases of learning based on optimization sequences with diversity. More details are referred to Section~\ref{sec:prediction}. 
    \item \textit{Intelligent searching} is to search for a desired optimization sequence of a target program through an improved particle swarm optimization algorithm~\add{\cite{kennedy1995particle,shi1999empirical,poli2007particle}}. In particular, we transform the compiler auto-tuning problem into a particle swarm optimization algorithm where an optimization sequence is regarded as a particle and then expand the search space by refining the parameters in the particle swarm optimization algorithm based on the performance similarity of generating sequences. More details are referred to Section~\ref{sec:pso}.
\end{itemize}

\subsection{\add{Performance-prediction model building}}
\label{sec:prediction}
To improve compiler auto-tuning efficiency, the proposed~\methodName~predicts the runtime performance of a compiled program with an optimization sequence, instead of actual execution. 

Intuitively, a prediction model with high accuracy is expected to be learnt with many instances (i.e., the actual runtime performance of a compiled program with various optimization sequences), but labelling so many instances (i.e., executing the compiled program with various optimization sequences) is undoubtedly time consuming. 
To address this issue, we propose a lightweight learning approach which builds a prediction model through 
several phases of learning: initial learning builds a prediction model with a small number of randomly generated instances and enhancement learning improves the accuracy of the prediction model through several phases of learning by using carefully selected instances (with performance diversity). More specifically, the instances used in this learning process are optimization sequences whose labels are actual runtime performance of the compiled program with the corresponding optimization sequences. \add{In particular, the features of the training data are compiler optimization flags, each of which is set to 0 or 1 representing whether the corresponding optimization flag is off or on. Note that we do not build a unified prediction model for all programs, but an individual prediction model for a program\footnote{\add{In the evaluation, for each program in the benchmark, we build a prediction model and search for its compiler optimization sequence.}}.} 

\subsubsection{Initial learning}
In initial learning, \methodName~builds a prediction model with a very small number of random generated data. In particular, \methodName~first randomly generates a small number of optimization sequences, and collects the runtime performance of the compiled program with each optimization sequence \add{(i.e., input data of the ``Initial Build'' operation in Figure 1)}, which are regarded as the training data of initial learning. 
Note that in our experiment the number of training data used in initial learning is set to $2$. Then \methodName~builds a prediction model based on these data with a supervised learning algorithm (i.e., random forest in this paper), as the random forest algorithm~\cite{chen2021efficient} combines the decisions from multiple decision trees and performs well on data with high-dimensional features while a compiler optimization sequence has around one hundred optimization flags. 

\subsubsection{Enhancement learning}
As initial learning builds a prediction model with a very small number of data, its prediction may be far from accurate. To address this problem, \methodName~further selects a few more optimization sequences with more runtime performance diversity, which are used to further enhance the accuracy of the prediction model through several rounds of learning process. Since \methodName~builds a prediction model with a small number of training data, it is a lightweight learning approach.

In enhancement learning, \methodName~randomly generates a set of candidate optimization sequences and selects optimization sequences with diverse runtime performance, which are fed to train a prediction model. For a target program with an optimization sequence, the multiple trees of the random forest model generated by \methodName~give various predicted values on its runtime performance, and thus \methodName~combines the predicted performance resulting from different trees following Bayesian optimization ~\cite{chen2021efficient,cavazos2006automatic,snoek2012practical,brochu2010tutorial,frazier2018tutorial,shahriari2015taking}. 
That is, \methodName~combines the mean and variance on these predicted values to get the overall performance (denoted as $EI$) of the optimization sequence as Formula~\ref{EI}. \add{The ``Select'' operation in Figure 1 is conducted based on the $EI$ value of each candidate sequence.}
    \begin{equation}\label{EI}
    \begin{split}
    EI(sequence) =  (f'-\mu(sequence))\Phi(Z) + \sigma(sequence)\phi(Z)
                \\ where~ Z = \frac{f' - \mu(sequence)}{\sigma(sequence)}
    \end{split}
    \end{equation}
where $f'$ is the minimum execution time of optimization sequences to build the prediction model. For any $sequence$, $\mu(sequence)$ and $\sigma(sequence)$ represent the mean and variance of the prediction results of the trees in the random forest model. $\Phi(Z)$ is the cumulative distribution and $\phi(Z)$ is the probability density of standard Gaussian distribution on sequences performance.

To improve the accuracy of the prediction model, in each cycle of enhancing learning, 
\methodName~selects optimization sequences with performance diversity, i.e., considering good/bad runtime performance and randomness. 
Following this intuition, among the set of randomly generated sequences, besides selecting the optimization sequence with largest $EI$ value, \methodName~checks the accuracy of the sequence.
If the accuracy does not reach the standard (set to 0.95 in the experiment), \methodName~selects another optimization sequence with small $EI$ values with high probability. 
In particular, \methodName~first sorts all generated sequences based on the descent order of their $EI$ values, and assigns a probability to each optimization sequence by guaranteeing that the optimization sequence with small $EI$ values has large probability. Finally, \methodName~randomly selects an optimization sequence among all generated sequences, which might be an optimization sequence with low performance~\cite{siegmund2015performance,pereira2020sampling,lipowski2012roulette}.

With the selected optimization sequences (including their actual runtime performance), \methodName~trains another prediction model, \add{shown by the ``Rebuild'' operation in Figure 1}.~\methodName~repeats this enhancement learning process until the prediction model is accurate enough. That is, \methodName~achieves the desired accuracy of the prediction model through initial learning and several phases of enhancement learning. In the evaluation, the termination criterion of the learning process (i.e., the desired prediction accuracy) is set to 0.96\footnote{For some target programs, this accuracy may be hard to achieve. In such cases, to avoid training a prediction model with too much data, we also stop the training process when the number of training data exceeds 50.}. With such a build-tune process, \methodName~constructs a more accurate prediction model and selects better starting data for the intelligent searching process.

\subsubsection{\add{Algorithm Details}}
\add{More details of this prediction model building algorithm are given by Algorithm~\ref{alg:model building}.}

\add{The algorithm of performance-prediction model building has input parameters, $IniSize$ representing the number of instances for the initial model, $Acc$  representing the prediction accuracy threshold, and $N$ representing the number of instances for the final model.
Lines 2-3 initialize two vectors $Seqs$ and $PerofSeqs$, which represent the instances used for prediction-model building. Lines 5-12 generate instances for the initial performance-prediction model by randomly generating 0-1 sequences. In particular, Line 6 obtains each sequence's performance by actual execution. Line 13 builds a prediction model with a random forest model. }

\add{Lines 15-39 tune the prediction model by selecting new instances. In particular, lines 17-24 select a new sequence to tune the prediction model. Lines 18-19 obtain the performance of each candidate sequence and select the sequence with the best performance. Line 18 uses the prediction model to predict the candidate sequences. Lines 25-35 select another sequence according to performance diversity. Line 25 obtains the accuracy of the selected sequence by comparing its actual performance and predicted performance. Lines 27-35 check the accuracy and decide whether to select a new sequence. Lines 28-31 select a new sequence by giving lower-performing sequences a larger probability of being selected. Lines 37-38 decide whether this algorithm stops by checking the average prediction accuracy for the selected sequences. Lines 36 retrains the prediction model by the new instances.}

\begin{algorithm}[h]  
  \caption{Performance-Prediction Model Building}  
  \label{alg:model building}  
  \begin{algorithmic}[1]
    \Procedure{Building}{$IniSize, Acc, N$}
        \State $Seqs \gets []$
        \State $PerofSeqs \gets []$ \\
        \algorithmiccomment{~Lines 5-13 build the initial performance prediction model}
        \While{$Sizeof(Seqs) < IniSize$}
            \State $seq \gets random(0,1)_{number\_of\_flags}$
            \If{$seq$ not in $Seqs$}
                \State $per \gets Performance(seq)$ 
                \State $Seqs.append(seq)$ 
                \State $PerofSeqs.append(per)$ 
            \EndIf
  	    \EndWhile
  	    \State $model \gets RandomForestModel(Seqs, PerofSeqs)$ \\
  	    \algorithmiccomment{~Lines 15-39 tune the performance prediction model}
  	    \While{$Sizeof(Seqs) < N$}\\
  	    \algorithmiccomment{~Lines 17-24 selects sequences with highest EI from candidates}
  	        \State $candidates \gets Set(random(0,1)_{number\_of\_flags})$
            \State $model.predict( candidates)$
            \State $seq \gets selectFromcandidatesByEI$
            \If{$seq$ not in $Seqs$}
                \State $per \gets Performance(seq)$ 
                \State $Seqs.append(seq)$ 
                \State $PerofSeqs.append(per)$ 
            \EndIf
            \State $acc \gets Accuracy(seq)$\\
            \algorithmiccomment{~Lines 27-35 selects sequences by performance distribution from candidates}
            \If{$acc < Acc$}
                \State $seq\_new \gets selectFromcandidatesByValueDistribution$ 
                \While{$seq\_new$ in $Seqs$}
                    \State $seq\_new \gets selectFromcandidatesByValueDistribution$ 
                \EndWhile
                \State $per \gets Performance(seq\_new)$ 
                \State $Seqs.append(seq\_new)$ 
                \State $PerofSeqs.append(per)$ 
            \EndIf
            \State $model \gets RandomForestModel(Seqs, PerofSeqs)$
            \If{$AverageAcc(Seqs) > Acc$}
               break
            \EndIf
  	    \EndWhile
  	\EndProcedure
  \end{algorithmic}  
\end{algorithm}

\subsection{\add{Intelligent searching of optimization flags}}
\label{sec:pso}
To obtain a desired compiler optimization sequence for the target program, \methodName~reuses the data \add{(i.e., which are obtained by the ``Transform'' operation in Figure 1)} in building the prediction model as the starting point of the following searching process, and designs a particle swarm optimization algorithm. To avoid the local optimum problem, we further expand the search space by refining the parameters of the particle swarm optimization algorithm, which influence the exploring speed, resulting in an improved particle swarm optimization algorithm. In the following paragraphs, we first introduce how we solve the compiler auto-tuning problem through a particle swarm optimization algorithm and then how to improve this algorithm.

\subsubsection{Particle swarm optimization}
The particle swarm optimization algorithm~\cite{kennedy1995particle,shi1999empirical,poli2007particle} is a representative search algorithm, which accomplishes an optimization process by simulating the foraging behavior of birds and is widely used due to its extraordinary performance in balancing the local and global optimum. In our improved particle swarm optimization algorithm, we regard each optimization sequence $O$ as a \textbf{particle} whose initial movement velocity vector is represented by $V_i$ and initial position vector (representing each optimization sequence's enable/disable choices) is represented by $X_i$, and its following movement velocity vector is updated based on Formula~\ref{update}.
    \begin{small}
    \begin{equation}
    \label{update}
    \begin{split}
    V_i^{(t+1)} =  \omega \times V_i^{(t)} + c_1  \times rand_1^{(t)}  \times (pBest_i^{(t)} - X_i^{(t)}) \\ + c_2  \times rand_2^{(t)}  \times (gBest^{(t)} - X_i^{(t)})
    \end{split}
    \end{equation}
    \end{small}

where $\omega$ is the inertia weight, regulating the search range of the solution space. $c_1, c_2 $ are acceleration constants, which regulate the step size of the exploration. An increase in $c_1$ indicates an enhanced local search, and an increase in $c_2$  indicates an enhanced global search. $pBest$ is the historical optimal position vector of each particle and $gBest$ is the global historical position vector of all particles. $rand_1$ and $rand_2$ are randomly generated values between 0 and 1 in each iteration, which perturb for the particle's velocity update. In particular, $X_i^{(t)}$ is the particle's position in the $t_{th}$ iteration and $V_i^{(t)}$ is the particle's movement velocity in the $t_{th}$ iteration. Through these variables, \methodName~updates each particle's movement velocity vector for the $t+1_{th}$ iteration, as well as the particles' position through the updated velocity vector, as shown in Formulae~\ref{tmp} and~\ref{updatex}.
    
\begin{equation}\label{tmp}
    s_{i,j}^{(t)} = \frac{1}{1+e^{-v_(i,j)^{(t)}}} \\
\end{equation}
\begin{equation}\label{updatex}
    x_{i,j}^{(t)} = \left\{
    \begin{aligned}
    1, rand() < s_{i,j}^{(t)} \\
    0, rand() \geq s_{i,j}^{(t)} \\
    \end{aligned}
    \right. \\
\end{equation}
    
Due to the position vector of the particle consisting of 0 and 1, \methodName~adjusts the position vector of each particle by introducing a temporary vector $S_i$. In the $t_{th}$ iteration, the values of the temporary vector $S_i^{(t)}$ of each particle are obtained by the updated movement velocity vector $S_i^{(t)}$ via the sigmoid function as Formula~\ref{tmp}, which turns the value range of a temporary vector to be $[0,1]$. 
    
Through the temporary vectors, \methodName~updates the position vector for each particle according to Formula~\ref{updatex}. For particle $i$, if the $j_{th}$ element $s_{i,j}$ of the temporary vector $S_{i}$ is greater than a random value (whose value range is $[0,1]$), the $j_{th}$ element $x_{i,j}$ of the position vector $X_i$ is updated to $1$; otherwise, it is updated to $0$. In this way, \methodName~completes the update for the velocity vector and the position vector of each particle in this iteration\add{, shown by the ``Evolve'' operation in Figure 1}.

In each iteration, \methodName~uses the prediction model (generated in Section~\ref{sec:prediction}) to predict the runtime performance (i.e., execution time in this paper) of a generated optimization sequence (i.e., a particle's position vector) rather than actual execution, to reduce the time spent during compiler auto-tuning, \add{shown by the ``Predict'' operation in Figure 1}. In particular, in each iteration \methodName~updates the optimal position $gBest_i$ of each particle and the optimal position $pBest$ of all particles by the predicted runtime performance of an optimization sequence (i.e., the position vector of each particle).
    
\subsubsection{Improvement on Particle Swarm Optimization}    
To avoid falling into local optimization, \methodName~further expands the search space of optimization sequences by using sequences with good performance and sequences with bad performance in different ways. In particular, \methodName~identifies the optimal sequence among all the sequences generated during each iteration based on their predicted performance, and classifies all these sequences into two sets (i.e., a set of sequences with good performance and a set of sequences with bad performance) based on whether a sequence achieves similar performance as the optimal sequence. Moreover, \methodName~decides whether an optimization sequence achieves similar performance as the optimal sequence based on their sequence similarity since the acceleration effect of similar optimization sequences also have certain proximity. In particular, \methodName~uses Formula~\ref{sim} (i.e., the cosine similarity of two vectors $X_i$ and $X_j$~\cite{ye2011cosine}) to calculate the similarity between a sequence (denote as $X_i$) and the optimal sequence (denote as $X_j$), and represents their $s_{th}$ compiler optimization flag (0 or 1) by $x_{i,s}$ and $x_{j,s}$.
    
\begin{equation}\label{sim}
    sim(X_i,X_j) =\frac{\sum^n_{s=1}(x_{i,s} \times x_{j,s})}{\sqrt{\sum_{s=1}^n(x_{i,s})^2} \times \sqrt{\sum_{t=1}^n(x_{j,s})^2}}
\end{equation}
    
Based on this formula, \methodName~classifies the generated sequences into a set of sequences with strong similarity with the optimal sequence (i.e., set of sequences with good performance, denoted as $part_1$), and a set of sequences with poor similarity with the optimal sequence (i.e., set of sequences with bad performance, denoted as $part_2$). Furthermore, if the optimal sequence's runtime performance generated during this iteration is better than existing optimal sequences' runtime performance, \methodName~increases the $c_1$ value for particles in $part_1$ to strengthen the local search, trying to find a better result among the current optimal value, and increases the $c_2$ value for particles in $part_2$ to strengthen the global search, trying to get the desired result in the bigger search space.

\begin{algorithm}[h]  
    \caption{Intelligent ~ Searching}
    \label{alg:intellsearch}
    \begin{algorithmic}[1]
        \Procedure{Searching}{$IniPars,seq_{initial},s_{initial}$}
        \State $bestPerofEach \gets PerofEachParofIniPars$
        \State $bestSeqofEach \gets SeqofEachParofIniPars$
        \State $bestPerofAll \gets s_{initial}$
        \State $bestSeqofAll \gets seq_{initial}$
        \State $velofEach \gets Randomly Generated$ \\
        \add{\algorithmiccomment{~Lines 8-9 update the particles to generate new sequences}}
        \For{$i \gets 1 ~ to ~\mathcal{M}$} 
            \If{$i = 1$}
                \State $NewPars \gets Evolution(IniPars, velofEach)$\\
            \add{\algorithmiccomment{~Lines 13-22 predict the particles' position vector and update each particles}}
            \Else
                \State $prePer \gets []$
                \For{$j \gets 1~to~sizeof(IniPars)$}
                    \State $prePer[j] \gets model.predict(NewPars[j])$
                
                \If{$prePer[j] \leq bestPerofEach[j]$}
                    \State $bestPerofEach[j] \gets prePer[j]$
                    \State $bestSeqofEach[j] \gets NewPars[j]$
                \Else
                    \State continue
                \EndIf
                \EndFor\\
                \add{\algorithmiccomment{~Lines 23-33 update the particles to generate new sequences}}
                \State $CurrentBestPer \gets best(bestPerofAll)$
                \If{$CurrentBestPer > bestPerfofAll$}
                    \State $bestPerofAll \gets CurrentBestPer$
                    \State $bestSeqofAll \gets best(bestSeqofEach)$ 
                    \State $NewPars \gets Evolution(NewPars)$
                \Else
                    \State $NewPars_1, NewPars_2 \leftarrow divide(NewPars)$
                    \State $NewPars_1 \leftarrow Evolution(NewPars_1)$
                    \State $NewPars_2 \leftarrow Evolution(NewPars_2)$
                    \State $NewPars \leftarrow Combine(NewPars_1, NewPars_2)$
                
                \EndIf
            \EndIf
        \EndFor
       \State $(seq_{final}, s_{final})  \gets (bestSeqofAll, bestPerofAll)$ 
       \State \textbf{return} $(seq_{final}, s_{final})$
       \EndProcedure
   \end{algorithmic}
\end{algorithm}

\subsubsection{Algorithm Details}

More details of this improved particle swarm optimization algorithm are given by Algorithm~\ref{alg:intellsearch}. 

\add{Lines 2-6 initialize particles for the searching process. Lines 2-5 initialize the optimal position vector (i.e., sequence) and performance (i.e., the execution time of the target program) of each particle by $InitialPars$ (i.e., the data set used to build a prediction model), and the optimal position vector and performance of all the particles by $seq_{initial}$ (i.e., the sequence with the best performance in building a prediction model) and $s_{initial}$ (i.e., the best performance in building a prediction model). Line 6 randomly generates a velocity vector for each particle.}

\add{Lines 7-37 perform an iterative update of the entire particle swarm, where the position vector of each particle continuously changes during searching (i.e., generating new sequences) and the new sequences are predicted by the generated prediction model. \methodName~continuously selects the sequence with the best performance until reaches the termination condition. 
Lines 9-10 update each particle's position vector in the first round of evolution by using randomly generated velocity vector according to Formulae ~\ref{tmp} and ~\ref{updatex}.
In the subsequent iterations, Lines 12-22 update the best position and performance of each particle. Lines 15 predicts the new position vector of each particle updated in the previous iteration, and then Lines 16-21 update the optimal position vector (i.e., compiler optimization sequence) and optimal performance of each particle.} 

\add{Lines 24-35 update the position vector of each particle again in the new evolution. To avoid the search process falling into a local optimum, \methodName~provides two updating ways. 
Line 24 first obtains the optimal performance of all particles in this iteration. If this performance is better than the previous optimal performance of all particles, Lines 25-28 update the optimal performance and the optimal position of all particles, as well as the particles' position vectors as Formulae~\ref{update}, ~\ref{tmp} and ~\ref{updatex}. Conversely, Lines 29-33 divide all the particles into two groups based on the optimal sequence of this iteration as Formula ~\ref{sim}, and update the two groups in different ways (i.e., adjusting the values of $c_1$ and $c_2$ in Formula~\ref{update}), and finally merge the particles in the two groups.
Lines 34-35 obtain the desired compiler sequence and the execution time of the compiled program under the optimization sequence.}

\section{EXPERIMENTAL SETUP}
\label{sec:setup}
To evaluate the performance of \methodName, this experiment is designed to answer two research questions.
\begin{itemize}
  \item \textbf{RQ1: Effectiveness.} How does \methodName~perform compared to the state-of-the-art compiler auto-tuning techniques?
  \item \textbf{RQ2: Ablation analysis.} How does each component of \methodName~contribute to the performance?
\end{itemize}

\subsection{Dataset}
In this experiment, we choose two widely used C compilers GCC 8.3.0~\cite{gccsource} and LLVM 9.0.0~\cite{llvmsource} as target compilers, \add{which have been used in numerous
studies\cite{du2020roofline, du2020research, engelke2020robust, qiu2020challenging}\footnote{\add{To verify the effectiveness of \methodName~on the latest compilers, we also experiment with GCC 12.2.0 and LLVM 15.0.0, and discuss their results in Section~\ref{sec:version}.}}.}
Like the state-of-the-art work does~\cite{chen2021efficient}, we use \add{the same} programs from two widely used benchmarks cBench~\cite{cbenchsource} and PolyBench~\cite{polybenchsource} as target programs.

Table~\ref{t1} presents the statistics of these compiled programs. The first ten programs are from cBench and others are from PolyBench. That is, in this experiment, we have 20 compiled programs that cover a wide range of practical functions, ranging from 200 to 26,000 lines of code.

\begin{table}[H]
\caption{Statistics of programs} \label{t1}
\setlength{\tabcolsep}{3mm}{
\begin{tabular}{l||l|l|r}
\toprule[2pt]
\textbf{ID} & \textbf{Program} & \textbf{Function} & \textbf{Lines} \\
\midrule
P1 & correlation & Compute correlation & 248 \\
P2 & covariance & Compute Covariance & 218 \\
P3 & symm & Symmetric Matrix multiplication & 231 \\
P4 & 2mm & 2 matrix multiplications & 252 \\
P5 & 3mm & 3 matrix multiplications & 267  \\
P6 & cholesky & Cholesky decomposition & 212 \\
P7 & lu & LU decomposition & 210 \\
P8 & nussinov & Predict RNA folding & 569 \\
P9 & heat-3d & Heat Equation over 3D space & 211 \\
P10 & jacobi-2d & Jacobi-style stencil computation & 200 \\
\midrule
C1 & automotive\_bitcount & Testing bit manipulation & 954 \\
C2 & automotive\_susan\_e & Edge image recognition & 2,129 \\
C3 & automotive\_susan\_c & Corner image recognition & 2,129 \\
C4 & automotive\_susan\_s & Image smoothing & 2,129 \\
C5 & consumer\_tiff2rgba & TIFF image conversion & 22,321 \\
C6 & consumer\_jpeg\_c & Image compression & 26,950 \\
C7 & office\_rsynth & Text to speech synthesis & 5,412 \\
C8 & security\_sha & Secure hash algorithm & 297\\
C9 & bzip2e & File compression & 7,200 \\
C10 & telecom\_adpcm\_c & Pulse Code Modulation & 389 \\
\bottomrule[2pt]
\end{tabular}}
\end{table}

\subsection{Compared Techniques}
The existing compiler auto-tuning techniques are classified into unsupervised learning based techniques and supervised learning based techniques~\cite{ashouri2018survey, ashouri2016compiler}. 

\textbf{Unsupervised leaning based techniques} select a compiler optimization sequence with good performance via a specific search strategy and actual execution of the target program in the optimization space. 
We consider two representative unsupervised leaning based techniques, random iterative optimization~\cite{chen2012deconstructing} (abbreviated as \textbf{RIO}, which is a simple unsupervised leaning-based technique) and genetic algorithm based technique~\cite{garciarena2016evolutionary} (abbreviated as \textbf{GA}, which is the state-of-art unsupervised leaning-based technique). \add{Besides, in this study we also consider \textbf{OpenTuner}\footnote{\add{ https://github.com/jansel/opentuner}}, which is a general tuning framework in Github, not specialized for compiler auto-tuning.} 

\textbf{Supervised leaning based techniques} learn performance knowledge from some existing optimization sequences, then generate new compiler optimization sequences using some methods and predict them, finally select an ideal compilation optimization sequence. 
We consider two representative supervised leaning-based techniques, \textbf{BOCA}~\cite{chen2021efficient} (which is the state-of-art compiler auto-tuning technique) and \textbf{TPE}~\cite{bergstra2011algorithms} (which is actually a general but state-of-art Bayesian optimization based technique, not specific for compiler auto-tuning).

To alleviate the influence of randomness resulting from the proposed technique and compared techniques, we repeat each technique five times. 

\subsection{Implementation}
We implement \methodName~in Python based on numpy~\cite{numpybenchsource} and scikit-learn~\cite{skbenchsource}. 
We adopt the implementation of Random Forest Model in scikit-learn with default parameter settings and set parameter values for the proposed particle swarm optimization algorithm based on some trials.
That is, the parameters $c_1$, $c_2$, and $w$ are set to be 2, 2, and 0.6.

In implementing the compared techniques, we directly use their code if they provide a reproducible package \add{(i.e., for techniques TPE, BOCA, and OpenTuner)}, otherwise, we re-implement the techniques \add{(i.e., GA and RIO)} strictly following the description in the papers.

The experiment is performed on a workstation with Intel(R) Xeon(R) Gold 5218R CPU @ 2.10GHz, 377.8G memory, Ubuntu 16.04.6 LTS operating system.

\subsection{Measurement}

As previous work did~\cite{chen2021efficient}, in this experiment, we measure a compiler auto-tuning technique's runtime performance improvement by comparing against the compiler's default optimization. 
In particular, GCC and LLVM have -O1/O2/O3\add{/Ofast} optimization settings 
and we use their -O3 optimization settings as the compiler's default optimization in the study since they are highest default optimization settings of these compilers \add{and widely used by existing compiler tuning work~\cite{chen2021efficient}.}

In order to obtain the runtime performance improvement of a compiler auto-tuning technique, 
for any C program $p$, we compile $p$ with the optimization sequence selected by a compiler auto-tuning technique and with the default compiler optimization -O3 separately, resulting in two compiled programs $p_r$ and $p_d$. Then we run the two programs $p_r$ and $p_d$, and calculate the \textit{runtime performance acceleration} \textbf{\add{speedup}} of $p$ (achieved by a compiler auto-tuning technique) through dividing the execution time of $p_r$ by the execution time of $p_d$. We use the result as evaluation metrics for compiler auto-tuning techniques. \add{We obtain the the execution time of $p_r$ and $p_d$ by the \textbf{time} command of linux.}

\section{RESULTS AND ANALYSIS}
\label{sec:results}
\subsection{Overall Effectiveness}
\label{se4-1}
\subsubsection{Results of different techniques}

\begin{table*}
\small
\caption{\add{Performance improvement for programs over -O3 on GCC}}
\label{t3}
\setlength{\tabcolsep}{0.5mm}{
\begin{tabular}{c||ccccccccccc}
\toprule[2pt]
Technique & ID & \add{Speedup} & ID & \add{Speedup} & ID & \add{Speedup} & ID & \add{Speedup} & ID & \add{Speedup}\\
\midrule
CompTuner & \multirow{5}{*}{P1} & \textbf{1.077(3107)}  & \multirow{5}{*}{P2} & \textbf{1.080(4067)}  & \multirow{5}{*}{P3} & 1.042(2573)  & \multirow{5}{*}{P4} & 1.071(3720) & \multirow{5}{*}{P5} & \textbf{1.041(2976)}\\
RIO & &\XSolidBold && \XSolidBold &&  1.042(4172)  && \XSolidBold && \XSolidBold\\
GA  & &\XSolidBold && \XSolidBold &&  \XSolidBold  && \XSolidBold && 1.041(3160) \\
TPE & &\XSolidBold && \XSolidBold &&  1.046(3775)  && \textbf{1.072(3112)} && \XSolidBold\\
BOCA& &\XSolidBold && \XSolidBold &&  \textbf{1.075(1923)} &&  1.071(3726) &&  1.046(3639)\\
\add{OpenTuner}& &\add{\XSolidBold}& &\add{\XSolidBold}& &\add{\XSolidBold}& &\add{1.075(4691)}& &\add{\XSolidBold}\\
\midrule
CompTuner & \multirow{5}{*}{P6} & 1.013(4726)  & \multirow{5}{*}{P7} & \textbf{1.073(5549)}  & \multirow{5}{*}{P8} & 1.029(3661)  & \multirow{5}{*}{P9} & 1.025(2976) & \multirow{5}{*}{P10} & \textbf{1.055(2192)}\\
RIO & &\textbf{1.016(3018)} && \XSolidBold &&  \textbf{1.029(3264)}  && \XSolidBold && \XSolidBold\\
GA  & &1.013(3862)  && \XSolidBold&&  \XSolidBold  && 1.025(3684) && \XSolidBold\\
TPE & &\XSolidBold && \XSolidBold &&  \XSolidBold  && \textbf{1.027(2637)} && \XSolidBold\\
BOCA& &1.014(4971) && \XSolidBold && 1.030(4082)  && 1.028(3420) && 1.055(3026)\\
\add{OpenTuner}& &\add{\XSolidBold}& &\add{1.075(6792)}& &\add{1.033(4970)
}& &\add{\XSolidBold}& &\add{\XSolidBold}\\
\midrule
CompTuner & \multirow{5}{*}{C1} & 1.382(4642)  & \multirow{5}{*}{C2} & 1.424(4905)  & \multirow{5}{*}{C3} & \textbf{1.327(3907)}  & \multirow{5}{*}{C4} & 1.118(3018) & \multirow{5}{*}{C5} & 1.036(5527)\\
RIO & &\XSolidBold && \XSolidBold &&  \XSolidBold  && 1.119(3976) && \XSolidBold\\
GA  & &\XSolidBold && \XSolidBold &&  \XSolidBold  && \XSolidBold && \textbf{1.038(5028)}\\
TPE & &\XSolidBold && \XSolidBold &&  1.327(4527)  && \textbf{1.142(2998)} && \XSolidBold\\
BOCA& &\textbf{1.397(3952)} && \textbf{1.498(4850)} &&  1.327(4382)  && 1.127(3192) && \XSolidBold \\
\add{OpenTuner}& &\add{\XSolidBold}& &\add{1.431(5782)}& &\add{\XSolidBold
}& &\add{1.224(4102)}& &\add{\XSolidBold}\\
\midrule
CompTuner & \multirow{5}{*}{C6} & 1.018(4869)  & \multirow{5}{*}{C7} & \textbf{1.451(5019)}  & \multirow{5}{*}{C8} & 1.038(4892) & \multirow{5}{*}{C9} & \textbf{1.020(3628)} & \multirow{5}{*}{C10} & 1.010(2324)\\
RIO & &\XSolidBold && \XSolidBold &&  \XSolidBold  && \XSolidBold && 1.011(2971)\\
GA  & &\textbf{1.018(3960)} && \XSolidBold &&  \XSolidBold  && \XSolidBold && \XSolidBold \\
TPE & &\XSolidBold && 1.452(5927) &&  \XSolidBold  && 1.023(3785) && 1.014(2672) \\
BOCA& &1.025(4027) && \XSolidBold &&  \textbf{1.040(4285)}  && \XSolidBold && \textbf{1.017(1852)} \\
\add{OpenTuner}& &\add{\XSolidBold}& &\add{\XSolidBold}& &\add{1.039(5562)
}& &\add{1.026(5179)}& &\add{\XSolidBold}\\
\bottomrule[2pt]
\end{tabular}}
\end{table*}

\begin{table*}  
\small
\caption{\add{Performance improvement for programs over -O3 on LLVM}}
\label{t4}
\setlength{\tabcolsep}{0.5mm}{
\begin{tabular}{c||ccccccccccc}
\toprule[2pt]
Technique & ID & \add{Speedup} & ID & \add{Speedup} & ID & \add{Speedup} & ID & \add{Speedup} & ID & \add{Speedup}\\
\midrule
CompTuner & \multirow{5}{*}{P1} & \textbf{1.042(2052)}   & \multirow{5}{*}{P2} & 1.025(2439)  & \multirow{5}{*}{P3} & 1.041(2182)  & \multirow{5}{*}{P4} & \textbf{1.078(3085)} & \multirow{5}{*}{P5} & \textbf{1.066(2994)}\\
RIO & &\XSolidBold && 1.027(2271) &&  \XSolidBold  && \XSolidBold && \XSolidBold\\
GA  & &\XSolidBold && \XSolidBold &&  1.041(2267)  && \XSolidBold && \XSolidBold\\
TPE & &\XSolidBold && \XSolidBold &&  1.063(2379)  && 1.079(4673)  && \XSolidBold\\
BOCA & & 1.042(2892) && 1.030(2189) &&  \textbf{1.043(2074)}  && 1.078(3562)  && \XSolidBold \\
\add{OpenTuner} & & \add{\XSolidBold} && \add{\textbf{1.025(2174)}} && \add{1.043(2991)} && \add{\XSolidBold} && \add{1.066(4157)}\\
\midrule
CompTuner & \multirow{5}{*}{P6} & 1.013(5963)  & \multirow{5}{*}{P7} & 1.023(5728)  & \multirow{5}{*}{P8} & \textbf{1.116(3625)}  & \multirow{5}{*}{P9} &1.014(2230) & \multirow{5}{*}{P10} & 1.036(2874)\\
RIO & &1.015(6124) && \textbf{1.040(4419)} &&  \XSolidBold  && \XSolidBold && \XSolidBold\\
GA  & &\XSolidBold && \XSolidBold &&  \XSolidBold  && \XSolidBold && 1.036(3627)\\
TPE & &\XSolidBold && \XSolidBold &&  \XSolidBold  && \XSolidBold && 1.038(3135) \\
BOCA & &\textbf{1.015(5732)}  && \XSolidBold && 1.124(3824)  && \textbf{1.015(2152)}  && \textbf{1.039(2610)} \\
\add{OpenTuner} & & \add{\XSolidBold} && \add{1.027(4993)} && \add{\XSolidBold} && \add{1.014(2962)} && \add{\XSolidBold}\\
\midrule
CompTuner & \multirow{5}{*}{C1} & \textbf{1.030(3327)}  & \multirow{5}{*}{C2} & 1.069(4126)  & \multirow{5}{*}{C3} & \textbf{1.204(3294)}  & \multirow{5}{*}{C4} & 1.099(2965) & \multirow{5}{*}{C5} & 1.019(5524)\\
RIO & &\XSolidBold &&\XSolidBold && \XSolidBold &&\XSolidBold && \XSolidBold\\ 
GA & &\XSolidBold && 1.069(4362) && \XSolidBold && 1.099(3642) && \textbf{1.019(3057)}\\
TPE & &\XSolidBold &&\XSolidBold  && 1.204(3894)   && \XSolidBold   && \XSolidBold  \\
BOCA & &\XSolidBold && \textbf{1.072(3920)} && 1.209(3362) && \XSolidBold &&\XSolidBold\\
\add{OpenTuner} & & \add{\XSolidBold} && \add{\XSolidBold} && \add{\XSolidBold} && \add{\textbf{1.108(2711)}} && \add{1.038(4717)}\\
\midrule
CompTuner & \multirow{5}{*}{C6} & \textbf{1.093(4170)} & \multirow{5}{*}{C7} & 1.025(2749) & \multirow{5}{*}{C8} & 1.020(3998) & \multirow{5}{*}{C9} & \textbf{1.036(3557)} & \multirow{5}{*}{C10} & 1.361(2758) \\
RIO & &\XSolidBold && \XSolidBold && 1.023(4872) && 1.038(3913) && 1.371(3724)\\
GA && \XSolidBold  && \XSolidBold && \XSolidBold && 1.044(3721) && \XSolidBold\\
TPE &&1.101(4535) && 1.025(3889) && \XSolidBold && 1.038(3659) && \textbf{1.370(2159)}\\
BOCA & & 1.093(5598) && \textbf{1.029(2597)} && \textbf{1.032(2892)} && \XSolidBold && 1.372(2352)\\
\add{OpenTuner}& & \add{\XSolidBold} && \add{1.028(3917)} && \add{1.022(4537)} && \add{\XSolidBold} && \add{\XSolidBold}\\
\bottomrule[2pt]
\end{tabular}}
\end{table*}

To investigate the optimization effect of our proposed \methodName~on the target programs, we compare \methodName~with
\add{five} state-of-art compiler auto-tuning techniques (i.e., RIO~\cite{chen2012deconstructing}, GA~\cite{garciarena2016evolutionary}, TPE~\cite{bergstra2011algorithms} BOCA~\cite{chen2021efficient} and \add{OpenTuner~\cite{DBLP:conf/IEEEpact/AnselKVRBOA14}}). Tables~\ref{t3} and~\ref{t4} present the comparison results. 

\add{The state-of-the-art work BOCA~\cite{chen2021efficient}~uses the iteration times as the termination condition (i.e., 60 iterations), but the iteration times does not match the tuning choice in practice}. To simulate practical usage of compiler auto-tuning, we set the time limitation of \methodName~to be 6,000 seconds and run \methodName~within 6,000 seconds for each target program, recording: (1) the runtime performance acceleration achieved by the generated optimization sequence (denoted as $performance_c$), and (2) how much time is used to generate this optimization sequence (denoted as $time_c$). The corresponding results are given by the \textbf{Speedup} values of \methodName~in the tables, where the values before the parentheses represent $performance_c$, and the values within the parentheses represent $time_c$. Note that $time_c$ is no larger than the given 6,000 seconds. \add{To get $time_c$, we record how much time \methodName~consumes by \textbf{time()} function of the Python library. If the time consumption reaches 6,000 seconds, we stop the tuning process. We record the optimal acceleration result and $time_c$ to achieve the result of this process. For the comparison technique, we repeat the tuning process with a time limit of $1.5 * time_c$.}
Then we explore the performance of the compared techniques based on whether they can generate a compiler optimization sequence with the same runtime performance acceleration $performance_c$. In particular, we run each compared technique within larger time limitation and record (1) whether any generated optimization sequence can reach the runtime performance acceleration $performance_c$, and (2) how much time is used to generate this optimization sequence. The corresponding results are given by the \textbf{Speedup} values of RIO, GA, TPE,
BOCA\add{, and OpenTuner} in the tables, where \XSolidBold~represents the compared technique does not generate an optimization sequence with such high runtime performance acceleration and the values within the parentheses represent the time used to generate the corresponding sequence. If the compared technique generate an optimization sequence with such high runtime performance acceleration, we present its runtime performance acceleration and the time.
As each technique is repeated five times to alleviate the influence of randomness, we present the median results in these tables. The complete results are given in the reproducible package. Moreover, smaller time in the table indicates the corresponding auto-tuning technique is more effective. Note that for each target program we use the bold font to emphasize the best result of all studied auto-tuning techniques. 

From Tables~\ref{t3} and~\ref{t4}, the runtime performance acceleration of \methodName~on GCC ranges from 1.010 to 1.451, and on LLVM ranges from 1.013 to 1.361, indicating that 
(1) the optimization sequences generated by \methodName~are always better than the default -O3 optimization given by GCC and LLVM, and 
(2) \methodName~always accelerates target programs through generated compiler optimization sequences but achieves different acceleration on different target programs. Besides, the time of \methodName~used to generate an optimization sequence with good runtime performance (i.e., the values within the parenthesis $time_c$) is usually smaller than the given time limitation 6,000 seconds, indicating that \methodName~can be improved in the future.

Many results of the compared baselines (i.e., RIO, GA, TPE, BOCA\add{, and OpenTuner}) are marked with \XSolidBold, indicating that these compared baselines do not generate an optimization sequence with the same runtime performance acceleration as \methodName~even if they are given with longer time limitation. Moreover, on \add{2} programs (i.e., P1, P2
) on GCC and 
\add{1} programs (i.e.,
C1) on LLVM, none of the baselines generates an optimization sequence with the same runtime performance acceleration as \methodName~given their longer time limitation. Among the 
\add{five} baselines, BOCA performs much better, because RIO, GA, TPE,
BOCA\add{, and OpenTuner} have 29, 29, 24, 
12\add{, and 24} \XSolidBold~according to this table, respectively. For a compared technique, more \XSolidBold represents the technique cannot achieve the optimization results obtained by \methodName~even in more given time on more cases.

\subsubsection{Statistics analysis}
To further investigate whether \methodName~outperforms the compared baselines, we statistically analyze the experimental results from two perspectives. First, we explore the best technique by analyzing how many cases a technique achieves the best optimization result. Second, we perform a series of statistics analysis (including significant analysis) on \methodName~and the state-of-art technique BOCA.

\textbf{Comparison among all techniques.} 
From Tables~\ref{t3} and~\ref{t4}, for each program we identify the compiler auto-tuning technique that achieves the best optimization result (which is addressed by the font bold). Then for each compiler auto-tuning technique we calculate the number of programs the corresponding technique achieves the best optimization results, which are given by Table~\ref{statistic}. In this table, the last three columns present the statistics results for GCC, LLVM, and both compilers. 

From this table, among the 20 target programs, our proposed \methodName~achieves the best optimization result on 8/
\add{8} programs of GCC/LLVM respectively, while the compared RIO, GA, TPE, BOCA\add{, and OpenTuner} achieve the best results on only 2/
\add{1}, 2/1, 3/1, 
5/7\add{, and 0/2} programs of GCC/LLVM, respectively. This observation confirms that \methodName~outperforms the baselines in generating an optimization sequence with good performance in most cases. 

To sum up, our proposed technique \methodName~generates optimization sequences better than the default -O3 optimization of GCC and LLVM. Moreover, in most cases it outperforms the compared compiler auto-tuning techniques, including the latest work BOCA, demonstrating that \methodName~is promising. 
\begin{table}[H]
\caption{Number of best performance results}\label{statistic}
\begin{tabular}{ccccccccccc}
\toprule[2pt]
Technique & GCC & LLVM & All\\
\midrule
CompTuner & \textbf{8} & \textbf{\add{8}} & \textbf{\add{16}}\\
RIO & 2 & \add{1} &  \add{3}\\
GA & 2 & 1 & 3\\
TPE & 3 & 1 & 4\\
BOCA & 5 & 7 & 12\\
\add{OpenTuner} & \add{0} & \add{2}& \add{2}\\
\bottomrule[2pt]
\end{tabular}
\label{statistic}
\end{table} 

\textbf{Comparison between \methodName~and BOCA.} 
Since BOCA is the most recent work on compiler auto-tuning, we conduct a thorough comparison between \methodName~and BOCA. 

From Tables~\ref{t3} and~\ref{t4}, among the 20 target programs our proposed \methodName~achieves better results than BOCA on 14/11 programs for compiler GCC/LLVM respectively, while BOCA only achieves better results on 6/9 programs for compiler GCC/LLVM respectively. Even when neither \methodName~nor BOCA achieves the best result, \methodName~often outperforms BOCA (e.g. P6 on GCC). That is, \methodName~outperforms BOCA in most programs. We further investigate the 25 
programs (14 programs for GCC and 11 programs for LLVM) where \methodName~outperforms BOCA, comparing the time used by \methodName~and BOCA when achieving the same speedup\footnote{We set the time to 6000s if BOCA does not achieve the same optimization result in fixed time.}. On average, \methodName~saves the 20.56\% time of the 14 programs for GCC and 24.87\% time for the 11 programs for LLVM. That is, on these programs, \methodName~uses much smaller time than BOCA to achieve the same optimization results, which also indicates the outperformance of \methodName~over BOCA. 

Furthermore, we perform significance analysis on the 40 cases in the Tables~\ref{t3} and~\ref{t4}\footnote{In the same way as before, if BOCA fails to achieve the optimization result as \methodName~within the fixed time, its time is set to 6000s.} to compare the acceleration effect of \methodName~and BOCA.  By performing Pair T-test on the time of the 40 cases, we get the p-value 0.002986, which shows that the optimization flag settings generated by \methodName~are significantly better than those by BOCA.

We manually explore the optimization flag settings of some programs recommended by \methodName~and BOCA to learn why the former outperforms the latter. In particular, for each program we apply \methodName~and BOCA five times, manually analyze their recommended optimization flag settings, and find that the five optimization flag settings recommended by \methodName~are close to a large extent while the five optimization flag settings recommended by BOCA are very different. For example, for program P7 on GCC, \methodName~always enables a set of 30 optimization flags in all five experiments, while BOCA always enables only a set of 9 optimization flags. In other words, the optimization sequences recommended by BOCA for P7 in five experiments are very different. In other words, BOCA does not perform well as \methodName~because its prediction model does not recommend impactful flags successfully.

\subsubsection{Optimization sequences of CompTuner}
\add{To further investigate whether \methodName~ selects suitable compiler optimization flags for a program, we conduct a small study on the optimization sequences produced by \methodName. 

\textbf{Optimization sequences for P1 and C1.} 

Here we take programs P1 and C1 as examples and manually analyze the optimization sequences produced by CompTuner\footnote{\add{We put the sequences selected by \methodName~for all programs on our website.}}.

For program P1, \methodName~selects 81 optimization flags, including 7 loop-related optimization flags, 6 branch-related optimization flags, and 68 flags for other program structures. In particular, we find \methodName~selects many loop-related flags (e.g., -floop-unroll-and-jam, -floop-interchange). This result is consistent with our expectation, because the program P1 consists of five double-loop structures and one triple-loop structure, which accounts for most of the program. Among the remaining unselected flags, \methodName~does not select the -fcrossjumping flag, 
as the official GCC documentation interprets this flag as ``The resulting code may or may not perform better than without cross-jumping''.

For program C1, \methodName~selects 85 optimization flags, including 3 loop-related flags, 9 branch-related flags and 73 flags for other program structures. Compared with P1, C1 has more branch structures, so \methodName~selects most branch-related flags for C1 (e.g., -fcse-follow-jumps, -fguess-branch-probability). Moreover, since C1 contains many calculations, \methodName~also selects the corresponding flags (e.g., -fipa-icf, -ftree-bit-ccp). Besides, \methodName~does not select the -fgcse flag for C1, as the official GCC documentation interprets this flag as ``may get better run-time performance if adding -fno-gcse''.

\textbf{Optimization sequence comparison.} 

It is also interesting to learn how good the optimization sequences produced by \methodName~are by comparing against the best optimization sequences (i.e., best optimization configurations), and thus we conduct a small study in this section. We can hardly conduct a study on the whole set of optimization flags because a GCC compiler contains more than 100
optimization flags\footnote{\add{The searching space is larger than $2^{100}$, and thus it is very costly to find the best optimization
configuration for even one program.}}. Alternatively, we select the first three programs of Polybench and Cbench, i.e., programs P1, P2, P3, C1, C2, C3, and construct compilers with a smaller number of optimization flags. That is, for each studied compiler (i.e., GCC and LLVM), we construct two sets of compiler optimization flags, each of which contains 10 randomly selected compiler optimization flags. The four sets of compiler optimization flags are shown by Table~\ref{flags}. Within a set of 10 optimization flags we find its best configuration through a search process and compare the best configuration against optimization sequences produced by \methodName. 

\begin{table}[H]
\caption{\add{Compilers with selected optimization flags}}
\label{flags}
\small
\begin{tabular}{cc}
\toprule[2pt]
Compiler & Flags\\
\midrule
GCC1 & \makecell[l]{-fssa-phiopt, -fsched-interblock, -ftree-loop-distribution,\\ -ftree-loop-vectorize, -fcse-follow-jumps, -fthread-jumps,\\ -finline-small-functions, -fipa-cp, -ftree-dce, -ftree-dse} \\
\midrule
GCC2 & \makecell[l]{-fdelayed-branch, -fdse, -fipa-reference,\\ -ftree-sink, -fipa-vrp, -fipa-bit-cp, \\ -floop-unroll-and-jam, -floop-interchange, -fpeel-loops, -fgcse} \\
\midrule
LLVM1 & \makecell[l]{-targetlibinfo, -tbaa, -basicaa,\\ -globalopt, -ipsccp, -deadargelim,\\
-instcombine, -basiccg, -prune-eh, -inline} \\
\midrule 
LLVM2 & \makecell[l]{-functionattrs, -argpromotion, -domtree,\\ -early-cse, -lazy-value-info, -jump-threading, \\-correlated-propagation, -simplifycfg, -instcombine, -tailcallelim} \\
\bottomrule[2pt]
\end{tabular}
\end{table}

\begin{table}
\caption{\add{Comparison of CompTuner and best configurations}}
\label{construct}
\small
\begin{tabular}{cccccc}
\toprule[2pt]
Program & Performance & Unnecessary Flags & Missing Flags & Accuracy\\
\midrule
& & GCC1 & &\\
\midrule
\midrule
P1 & 0.953 & -fsched-interblock & -fthread-jumps & 0.8\\
P2 & 0.936 & -fcse-follow-jumps & \makecell[c]{-finline-small-functions, \\-ftree-dse} & 0.7\\
P3 & 0.964 & - & \makecell[c]{-fsched-interblock, \\-ftree-loop-vectorize} & 0.8 \\
C1 & 0.981 & -ftree-dce & - & 0.9\\
C2 & 0.905 & -fssa-phiopt & \makecell[c]{-finline-small-functions, \\-fipa-cp} & 0.7 \\
C3 & 0.947 & -ftree-dse & -ftree-loop-distribution & 0.8 \\
\midrule
& &  GCC2 & &\\
\midrule
\midrule
P1 & 0.966 & - & -ftree-sink & 0.9 \\
P2 & 0.949 & -fdse & -floop-interchange & 0.8 \\
P3 & 0.922 & -ftree-sink & -fipa-reference, -fgcse & 0.7\\
C1 & 0.913 & -fipa-bit-cp, fipa-vrp & -fgcse & 0.7\\
C2 & 0.935 & -fipa-bit-cp & -fpeel-loops & 0.8 \\
C3 & 0.971 & -fdse, -floop-unroll-and-jam & - & 0.8\\
\midrule
& & LLVM1 & &\\
\midrule
\midrule
P1 & 1.000 & - & - & 1.0\\
P2 & 0.976 & -tbaa & - & 0.9 \\
P3 & 0.947 & -basicaa & -ipsccp  & 0.8 \\
C1 & 0.980 & - & -ipsccp & 0.9 \\
C2 & 0.977 & - & -basiccg, -prune-eh  & 0.8 \\
C3 & 0.935 & -deadargelim & -tbaa & 0.8 \\
\midrule
& & LLVM2 & &\\
\midrule 
\midrule 
P1 & 0.984 & -early-cse & - & 0.9\\
P2 & 0.962 & -domtree & -jump-threading & 0.8\\
P3 & 0.973 & -instcombine, & -lazy-value-info & 0.8 \\
C1 & 0.991 & - & -functionattrs & 0.9 \\
C2 & 0.946 & -domtree & -jump-threading, -simplifycfg & 0.7 \\
C3 & 0.959 & -argpromotion & -correlated-propagation & 0.8 \\
\bottomrule[2pt]
\end{tabular}
\end{table}

Due to space limitation, we do not present the specific optimization sequences selected by \methodName~and the best configurations in the paper, but on the website. Alternatively, we analyze the difference between their selected optimization sequences and the difference on runtime-performance acceleration. The analysis results are given by Table~\ref{construct}, where the second column presents the ratio of the runtime performance of the program compiled with the optimization sequences selected by \methodName~to the best configurations, the third column presents the flags selected by \methodName, while not selected by the best configurations, the fourth column presents the flags selected by the best configurations, while not selected by \methodName, and the last column presents the ratio of optimization flags whose setting are the same for \methodName~and best configurations.

From the second column, the performance of \methodName~is usually close to 1, indicating that the optimization flags selected by \methodName~achieve close runtime performance acceleration as the best configurations. From the last column, the accuracy is always larger than 0.7 and mostly larger than 0.8, indicating that the optimization configurations recommended by \methodName~are very close to the best configurations. This conclusion can also be drawn by the third and fourth columns, i.e., \methodName~selects only at most 2 unnecessary flags or misses at most 2 flags.

Besides, we also observe that the runtime performance acceleration of \methodName~(i.e., given by the second column) usually increases as its accuracy (i.e., given by the last column). This observation is as expected because optimization sequences with high accuracy tends to have much overlap with the best configurations. Besides, there are some exception on the comparison between P1 and C3 on GCC2, and we suspect the reason to be that some optimization flag (e.g., -ftree-sink) may affect the runtime performance to a large extent than other flags.

}

\subsection{Ablation Analysis}
In this section, we perform an ablation analysis to investigate the contribution of two components of \methodName, which are a prediction model (given by Section~\ref{sec:prediction}) and intelligent searching process (given by Section~\ref{sec:pso}). 

\begin{figure}[htp]
\centering
    \begin{minipage}[t]{0.49\textwidth}
        \centering
        \includegraphics[width=1\textwidth]{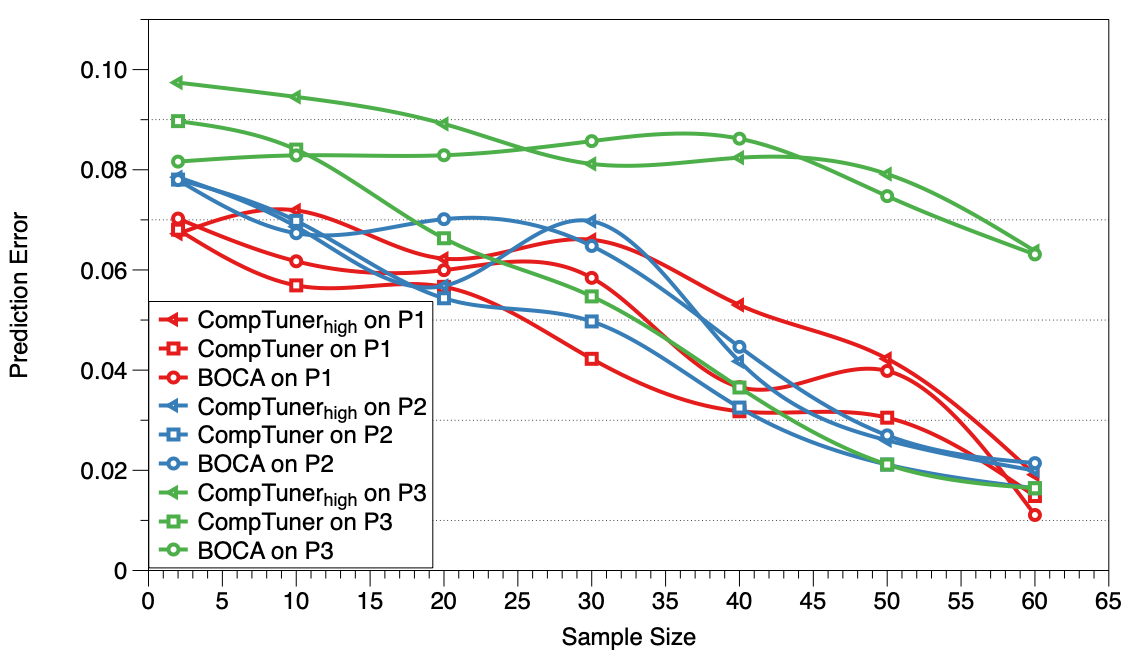}
        \caption{Prediction Error Rate of PolyBench}
        \label{ac1}
    \end{minipage}
    \begin{minipage}[t]{0.49\textwidth}
        \centering
        \includegraphics[width=1\textwidth]{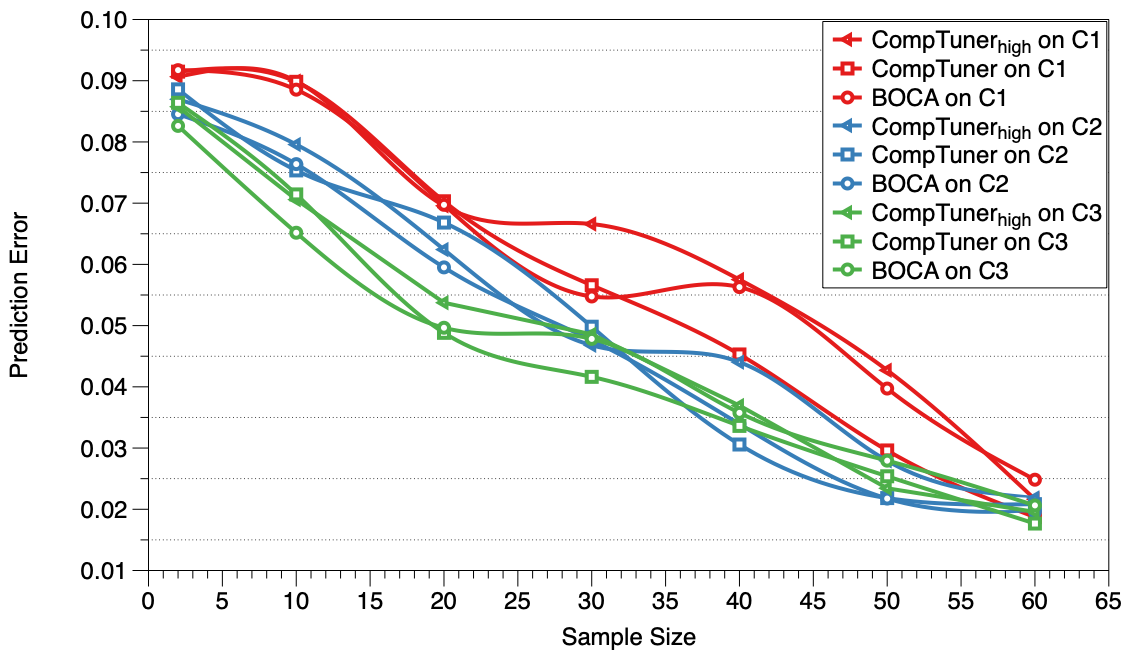}
        \caption{Prediction Error Rate of cBench}
        \label{ac2}
        \end{minipage}
\end{figure}
\subsubsection{Prediction model analysis}
To investigate whether the proposed lightweight learning approach in \methodName~is well designed, similar to BOCA, we build a prediction model by using only optimization sequences with good runtime performance, resulting in a variant of \methodName~(donated as $CompTuner_{high}$). Besides, BOCA is the state-of-art supervised learning based technique that also builds a prediction model. \add{However, different from \methodName, BOCA builds a prediction model by selecting the optimization sequences with the highest \textbf{expected improvement}}, and thus in this study we also compare against the prediction model built by BOCA. To sum up, in this study we compare the prediction models of \methodName, $CompTuner_{high}$, and BOCA, in terms of their prediction accuracy and runtime performance acceleration. Besides, all these approaches start building a prediction model with some randomly selected optimization sequences. To alleviate the influence of this random selection, we use the same randomly selected optimization sequences as the input of the three approaches, and observe the prediction accuracy of each compared technique by gradually selecting more optimization sequences in building the prediction model.

For each target program we randomly generate 100 compiler optimization sequences and record their actual runtime performance as the ground truth. For each compared technique (i.e., \methodName~, $CompTuner_{high}$, and BOCA), we first construct prediction models with a various numbers of training data (i.e., the number is set to be 2, 10, 20, 30, 40, 50 and 60), and use these prediction models to predict the runtime performance of the randomly generated optimization sequences, recording the accuracy of each prediction model. The results are given by Figures~\ref{ac1} and ~\ref{ac2}, where the horizontal axis represents the size of the training set, and the vertical axis represents the prediction error rate, i.e., the absolute difference between predicted and actual runtime performance divided by the actual runtime performance. A small prediction error rate means a good prediction model. In particular, this ablation study is conducted on GCC by using the first three target programs from PolyBench and cBench as the representatives, which are P1, P2, P3, and C1, C2, C3. To distinguish these programs, in the figure we use red to denote P1 and C1, blue to denote P2 and C2, and green to denote P3 and C3.

From these figures, although these techniques take the same optimization sequences as input to build a prediction model, their corresponding models have different prediction error rates. The prediction error rate of \methodName~is smaller than that of $CompTuner_{high}$ and BOCA in most programs  (i.e., four of the six programs when the input size is no smaller than 20). In particular, the prediction error rate of \methodName~on P3 is only 0.02 while those of $CompTuner_{high}$ and BOCA are larger than 0.06. Although sometimes (e.g., in P1 and P3) the prediction error rate of \methodName~is slightly larger than $CompTuner_{high}$ or BOCA at the very beginning of model construction (i.e., fed with a very small number of optimization sequences), the former becomes close to the latter when 20 optimization sequences are used. Moreover, the prediction error rate of \methodName~is usually smaller than $CompTuner_{high}$ and BOCA when more than 40 input data are used, which demonstrates \methodName~can always produce a prediction model with high accuracy given a reasonable number of training data.

Moreover, compared with $CompTuner_{high}$, the curves of \methodName~are close, indicating that the prediction accuracy of its models on various target programs is stable. On the other hand, from these figures we can observe that the prediction error rate of the prediction model constructed using either \methodName, BOCA or $CompTuner_{high}$ decreases when we use more training data. However when BOCA and $CompTuner_{high}$ use more training data to tune the model, the prediction error rate will increase (e.g., $CompTuner_{high}$ on P2, BOCA on P3), which indicates that these two techniques lack the ensure of the accuracy of the model when selecting more data to tune the model.

Then we compare the runtime performance acceleration of \methodName~and $CompTuner_{high}$ (whose only difference lies in how to build a prediction model) following the setup of Section~\ref{se4-1}. Note that we do not compare the runtime performance acceleration of \methodName~and BOCA because we have done so in Section~\ref{se4-1}. The comparison results are given by Row \methodName~as well as Row $CompTuner_{high}$ in Table \ref{ablation}. Compared to $CompTuner_{high}$, \methodName~achieves better optimization results on five of the six target programs, with an average improvement of 13.29\%\footnote{This percentage is calculated based on the average results of Table~\ref{ablation} by considering the five programs.}. In particular, the improvement of \methodName~over $CompTuner_{high}$ on P2 is 33.17\%. These comparative results reveal that the way of constructing the model by \methodName~not only constructs a more accurate prediction model, but also produces a better acceleration effect on the programs.

\begin{table}[H]
\caption{Results of the ablation study}\label{ablation}
\begin{tabular}{cccc}
\toprule[2pt]
Technique & P1 & P2 & P3  \\
\midrule
CompTuner & \textbf{1.052}  & 1.059 &  \textbf{1.029} \\
$CompTuner_{high}$ & 1.039 & 1.025  &  1.027\\
$CompTuner_{impact}$ & 1.051 & \textbf{1.067}  & 1.023 \\
\midrule
Technique & C1 & C2 & C3  \\
\midrule
CompTuner & \textbf{1.299}  & 1.365 & \textbf{1.274} \\
$CompTuner_{high}$ & 1.291 &  \textbf{1.416}   &  1.266\\
$CompTuner_{impact}$ & 1.286 & 1.393 & 1.259 \\
\bottomrule[2pt]
\end{tabular}
\end{table}

\subsubsection{Intelligent searching analysis}
To investigate whether the improved particle swarm optimization algorithm in \methodName~is well designed, we replace this optimization selection process with that of BOCA, which generates several compiler optimization sequences using all combinations of impactful compiler optimization flags and some combinations of less-impactful compiler optimization flags, resulting in a variant of \methodName~(donated as $CompTuner_{impact}$). The comparison results are also given in Table~\ref{ablation}. 

From this table, \methodName~achieves better optimization results on most programs (i.e., four of the six target programs), confirming the effectiveness of intelligent searching in \methodName. We also note that $CompTuner_{impact}$ achieves better optimization results than \methodName~on program P2, the smallest program of the six target programs, and thus we hypothesize that the performance of \methodName~is more obvious on larger programs. That is, combining impactful flags with our \methodName~may be a promising direction in the future.

\add{To verify our hypothesis, we choose some smaller programs (i.e., P1, P2, and P3) and larger programs (i.e., C4 and C5) and compare their performance of \methodName~and $CompTuner_{impact}$, as shown in Table~\ref{sus}. From this table, the advantage of \methodName~over $CompTuner_{impact}$ is more obvious on larger programs. 
In particular, P1, P2, and P3 represent programs with 200+ lines, while C4 and C5 represent programs with 2000+ lines. According to the experimental results, on the selected 
five programs, $CompTuner_{impact}$ outperforms \methodName~only on the small program P2 (with 218 lines). Moreover, \methodName~performs more prominent on the large program C5 (with 22,321 lines).}

\begin{table}[H]
\caption{\add{Comparison of \methodName~and $CompTuner_{impact}$}}
\label{sus}
\begin{tabular}{cccccc}
\toprule[2pt]
Technique & P1 & P2 & P3 & C4 & C5\\
\midrule
CompTuner & \textbf{1.052}  & 1.059 &  \textbf{1.029} &  \textbf{1.116} &  \textbf{1.035}\\
$CompTuner_{impact}$ & 1.051 & \textbf{1.067}  & 1.023 & 1.109 & 1.021\\
Difference & 0.001 & -0.008 & 0.006 & 0.070 & 0.014 \\
\bottomrule[2pt]
\end{tabular}
\end{table}

\section{Threats to Validity}
\label{sec:threats}
The internal threat mainly comes from the implementation of compiler auto-tuning approaches. To reduce this threat, we use the reproducible package of the compared approaches and re-implement the compared approaches strictly following their papers if their reproducible package is not available. Besides, the authors review the implementation to further reduce this threat.

The external threat mainly comes from the target programs and compilers. To reduce this threat, we use the same benchmarks and popular compilers as previous work did~\cite{chen2021efficient,garciarena2016evolutionary}. 

The construct threat lies in the metrics used in the evaluation. \add{GCC
and LLVM have several recommended optimization settings such as -O1, -O2, -O3, and -Ofast. -Ofast is reported to be an addition to -O3 with some unconventional optimizations that are achieved by breaking some international standards. -O1 or -O2 does not enable many optimization flags. Therefore, in the literature -Ofast is not generally recommended and the existing work on compiler tuning is usually compared against -O3. 
} As prior work did~\cite{chen2021efficient}, we also use the runtime performance acceleration over the compiler's default -O3 optimization. 

\section{Discussion}
\label{sec:discuss}
\add{In this section, we conduct three small studies to investigate the influence of compiler auto-tuning time, compiler versions, and multiple-phase model construction.}

\add{\subsection{Influence of the tuning time}}
\label{sec:time}
\add{Compiler auto-tuning time may influence the performance of auto-tuning techniques, and thus we compare the performance of compiler auto-tuning techniques by changing their tuning time limitation. In particular, we repeat the experiment in Section~\ref{se4-1} on only 6 programs (i.e., P1, P2, P3, C1, C2, and C3) by setting the time limitation to be 2,000 seconds, while the time limitation used in Section~\ref{se4-1} is 6,000 seconds. The results are given by Table~\ref{timechange}. }

\begin{table}[H]
\small
\caption{\add{Performance improvement for programs over -O3 on GCC and LLVM in 2,000 seconds}}
\label{timechange}
\setlength{\tabcolsep}{0.5mm}{
\begin{tabular}{ccccccc}
\toprule[2pt]
Technique & P1 & P2 & P3 & C1 & C2 & C3\\
\midrule
& & &  GCC  & & &\\
\midrule
\midrule
CompTuner & \textbf{1.051(1923)}  & \textbf{1.047(1662)} &  1.041(1825) & 1.215(1520) & 1.197(1394) & \textbf{1.263(1882)}\\
RIO & \XSolidBold & \XSolidBold & \XSolidBold & \XSolidBold & \textbf{1.197(1215)} & \XSolidBold\\ 
GA & \XSolidBold & \XSolidBold & 1.041(1858) & \XSolidBold & \XSolidBold & \XSolidBold\\
TPE & \XSolidBold & \XSolidBold & \XSolidBold & 1.240(1637)
& 1.199(1558) & 1.280(1904)\\
BOCA & \XSolidBold & 1.049(1780) & \textbf{1.046(1532)} & \textbf{1.264(1482)} & \XSolidBold & 1.276(1964)\\
OpenTuner & \XSolidBold & \XSolidBold & \XSolidBold & \XSolidBold & \XSolidBold & \XSolidBold\\
\midrule
& & &  LLVM  & & &\\
\midrule
\midrule
CompTuner & \textbf{1.038(1725)}  & \textbf{1.022(1554)} &  1.037(1475) & \textbf{1.020(1632)} & 1.046(1913) & \textbf{1.103(1844)}\\
RIO & \XSolidBold & \XSolidBold & 1.038(1926) & \XSolidBold & \XSolidBold & \XSolidBold\\ 
GA & \XSolidBold & 1.024(1824) & \XSolidBold & 1.021(1849) & \XSolidBold & \XSolidBold\\
TPE & \XSolidBold & 1.040(1998) & \textbf{1.037(1216)} & \XSolidBold & \XSolidBold & \XSolidBold\\
BOCA & \XSolidBold & 1.023(1794) &\XSolidBold & 1.022(1962) & \textbf{1.052(1565)} & \XSolidBold\\
OpenTuner & \XSolidBold & \XSolidBold & \XSolidBold & \XSolidBold & 1.046(1924) & \XSolidBold\\
\bottomrule[2pt]
\end{tabular}}
\end{table}

\add{From this table, \methodName~outperforms the compared techniques in most programs within 2,000 second evaluation. In particular, \methodName~achieves the best optimization result on 3/4 programs of GCC/LLVM respectively, while the compared RIO, GA, TPE,
and BOCA, and OpenTuner achieve the best results on only 1/0, 0/0, 0/1, 2/1 and 0/0 programs of GCC/LLVM. Compared with results within 6,000 seconds (given by Table~\ref{t3} and Table~\ref{t4}), the performance improvement of these auto-tuning techniques increases from 2,000 seconds to 6,000 seconds. This observation is as expected because longer tuning time may increase the model's performance. However, the advantageous of \methodName~over the compared techniques holds from 2,000 seconds to 6,000 seconds. } 


\add{\subsection{Influence of the compiler version}}
\label{sec:version}
\add{To reduce the threat to validity resulting from compiler versions, we further evaluate \methodName~by comparing against BOCA~\cite{chen2021efficient} on the latest versions of GCC and LLVM, i.e., GCC 12.2.0 and LLVM 15.0.0, using the same programs as Section~\ref{sec:time}. The results are given by Table~\ref{latest}.}

\begin{table}[H]
\caption{\add{Comparison of \methodName~and BOCA on GCC 12.2.0 and LLVM 15.0.0}}
\small
\label{latest}
\setlength{\tabcolsep}{0.5mm}{
\begin{tabular}{ccccccc}
\toprule[2pt]
Technique & P1 & P2 & P3 & C1 & C2 & C3\\
\midrule
& & &  GCC 12.2.0 & & &\\
\midrule
\midrule
CompTuner & \textbf{1.014(4517)}  & \textbf{1.028(5262)
} &  \textbf{1.008(3884)} &  \textbf{1.122(5753)} &  1.095(4031) & \textbf{1.169(3983)}\\
BOCA & \XSolidBold & \XSolidBold & 1.014(5105) & 1.126(6814) & \textbf{1.174(3372)} & \XSolidBold\\
\midrule
& & &  LLVM 15.0.0 & & &\\
\midrule
\midrule
CompTuner & \textbf{1.021(4362)}  & \textbf{1.003(3527)} &  \textbf{1.032(3380)} &  \textbf{1.025(4171)} & \textbf{1.041(5514)} & 1.096(4793)\\
BOCA & \XSolidBold & 1.016(4107) & \XSolidBold & \XSolidBold & \XSolidBold & \textbf{1.114(4623)}\\
\bottomrule[2pt]
\end{tabular}}
\end{table}

\add{From this table, all the data of \methodName~are larger than 1, indicating that our proposed \methodName~achieves better results than -O3 optimization on the latest versions of GCC and LLVM. \methodName~achieves the best optimization result on 5/5 programs of GCC 12.2.0/LLVM 15.0.0 respectively, while the compared BOCA achieves the best results on only 1/1 programs of GCC 12.2.0/LLVM 15.0.0. Note that for P3 on GCC and P2 on LLVM, we mark the results of \methodName~as the best result (addressed in bold font) because BOCA spends more time to achieve the same runtime performance acceleration as \methodName. In other words, BOCA cannot produce an optimization sequence in GCC with the acceleration 1.008 within 3,884 seconds on P3, neither in LVVM with 1.003 within 3,527 seconds on P2. 

To sum up, on the latest version of GCC and LLVM, our proposed \methodName~still outperforms the state-of-the-art BOCA on most programs. Moreover, combined with the observations on old versions of GCC and LLVM (i.e., results from Table~\ref{t3} and Table~\ref{t4}), \methodName~performs much better than BOCA on the latest version compared with the old version of the compilers.}

\add{\subsection{Influence of multiple-phase model construction}}
\label{sec:bulding}
\add{Our approach \methodName~builds a prediction model through multiple-phase learning instead of single-phase learning, with the purpose of improved prediction accuracy. To verify whether multi-phase model construction performs better than single-phase learning, we implement a variant of \methodName~by replacing its multiple-phase learning by one-phase learning, which is called $CompTuner_{one-time}$ and compare the performance of \methodName~and $CompTuner_{one-time}$ on the same 6 programs as Section~\ref{sec:time}. For fair comparison, $CompTuner_{one-time}$ uses 50 optimization sequences randomly selected from candidate sequences to build the prediction model,
while CompTuner uses 50 another optimization sequences selected by multi-phase selection to build the model. The results are given in Table~\ref{one-time}.

To verity the effectiveness of the performance-prediction model building part of our technique, we compare the performance of our technique with $CompTuner_{one-time}$, which constructs the performance-prediction model by one-phase selecting from candidate sequences.\footnote{\add{We all select 50 samples for the two performance-prediction model construction ways.}}. We choose 6 programs on GCC to compare the two techniques.

}
\begin{table}[H]
\caption{\add{Comparison of \methodName~and $CompTuner_{one-time}$}}
\small
\label{one-time}
\setlength{\tabcolsep}{0.5mm}{
\begin{tabular}{ccccccc}
\toprule[2pt]
Technique & P1 & P2 & P3 & C1 & C2 & C3 \\
\midrule
& & & Prediction error rate & & &\\
\midrule
\midrule
CompTuner & \textbf{0.031}  & \textbf{0.024} &  0.041  & \textbf{0.030}  & 0.032 & \textbf{0.025} \\
$CompTuner_{one-time}$ & 0.039 & 0.026 & \textbf{0.035} & 0.044 & \textbf{0.014} & 0.031\\
\midrule
& & & Speedup & & &\\
\midrule
\midrule
Technique & P1 & P2 & P3 & C1 & C2 & C3 \\
\midrule
CompTuner & \textbf{1.047}  & \textbf{1.053} &  \textbf{1.026}  & \textbf{1.285}  & 1.348 & \textbf{1.257} \\
$CompTuner_{one-time}$ & 1.028 & 1.034  &  1.014 & 1.246 &  \textbf{1.437}   & 1.201\\
\bottomrule[2pt]
\end{tabular}}
\end{table}

\add{From Table ~\ref{one-time}, \methodName~outperforms $CompTuner_{one-time}$ on 5/6 programs and constructs a more accurate prediction model on 4/6 programs. 
Moreover, a prediction model with low error rate, usually results in optimization sequences with good runtime performance. However, on program P3, $CompTuner_{one-time}$ constructs a prediction model with high accuracy, while its runtime performance is lower than \methodName. We suspect the reason to be that \methodName~chooses a better starting point for the improved particle swarm process.}

\section{RELATED WORK}
\label{sec:related}
This paper targets compiler auto-tuning, and thus is most related to compiler auto-tuning. Besides, this paper is also related to software tuning in general. Therefore, in this section, we will first give a quick review on the existing work on software tuning in Section~\ref{related:tuning} and then compiler auto-tuning in Section~\ref{related:compiler}.
\subsection{Software Tuning}
\label{related:tuning}
Tuning is an important problem in both engineering and research. Many systems and models often are set with a very large number of configurable parameters, which can be tuned to satisfy different demands. For example, data management systems like MySQL have parameters whose values can be tuned to improve performance and ensure security, web service systems like Ngnix have parameters whose values can be tuned to ensure their performance and stability~\cite{DBLP:conf/icse/HeJLYZ0WL22}, embedded systems have parameters whose value combination can be tuned to consume less energy at runtime~\cite{DBLP:conf/iccd/VazquezGS19, DBLP:conf/iccma/RuanguraiS19}, deep learning models have a lot of hyperparameters, which can be tuned to make the models obtain better performance (e.g., high prediction accuracy)~\cite{DBLP:journals/candc/NematzadehKTA22, DBLP:journals/corr/abs-2102-02711}, some software systems have many parameters whose values can be tuned to achieve better runtime performance in different hardware environments~\cite{DBLP:conf/amcc/0038JL0LHML21}. 

Tuning occurs in various software and systems, but their common difficulty lies in how to efficiently explore the huge parameter space. To address this problem, researchers proposed many tuning techniques, which can be classified into search-based techniques and learning-based techniques. In particular, besides random based techniques~\cite{chen2012deconstructing} and genetic algorithm based techniques~\cite{garciarena2016evolutionary,hoste2008cole}, simulated annealing algorithm based techniques~\cite{DBLP:journals/corr/abs-1906-01504} were widely used, which found an optimal parameter setting through the actual execution of the target system and a specific search strategy. However, due to software complexity, actual execution consumes a lot of time and computing resources, resulting in low optimization efficiency. The learning based techniques build prediction models through some data and then predict the performance of parameter combinations instead of actual execution. To find better parameter combinations quickly, many sampling techniques were proposed to approximate the optimal parameter combinations in the large parameter space. For example, Henard et al.~\cite{7194602}~proposed a multi-objective optimization technique to reduce the search space by constraints, Nair et al.~\cite{8469102} and Victoria et al.~\cite{article} proposed Bayesian optimization based techniques to produce optimal samples through evaluating the performance of the samples by acquisition functions, Dogru et al.~\cite{DBLP:journals/cce/DogruVIWSHXNB22} proposed a reinforcement learning based technique, which uses a reward function to obtain the future benefits of the sampling to gradually select the parameter combinations in the search space. \add{Besides, Chris et al. ~\cite{DBLP:conf/IEEEpact/CumminsP0L17} proposed a deep neural network , which learns heuristics over raw code, constructs appropriate representations of the code and learns how best to optimize.}

\subsection{Compiler Auto-tuning}
\label{related:compiler}
\add{Compiler auto-tuning aims at recommending optimization settings to improve program runtime performance. Typically, compiler auto-tuning consists of flag selection and phase ordering. Phase ordering aims at deciding the order in which the selected optimization flags are applied. For example, Ashouri et al.~\cite{ashouri2017micomp} proposed a learning-based technique, which clusters the optimization passes of LLVM into different clusters to predict the speedup of a complete sequence by a pre-train prediction model. Huang et al.~\cite{huang2019autophase} proposed a reinforcement learning based technique, which predicts the best next flag by features of the program and previously applied passes. As this paper focuses on flag selection instead of phase ordering, we classify the existing techniques on flag selection~\cite{ashouri2018survey, ashouri2016compiler} into unsupervised learning based methods and supervised learning based methods, and review the existing techniques following this classification to review.}

\textbf{Unsupervised learning based techniques.} The unsupervised learning based techniques start the optimization process from randomly generated samples and use some searching strategies to generate new optimization sequences. Chen et al. ~\cite{chen2012deconstructing} proposed a random iterative optimization technique, which aims to generate a desired optimization sequence by generating new sequences in a random way and is usually regarded as the baseline for compiler auto-tuning techniques. To trace the searching process, some techniques (e.g., hill-climbing-based techniques~\cite{almagor2004finding} and genetic algorithm based techniques~\cite{garciarena2016evolutionary,hoste2008cole,ni2019evolutionary,sandran2012optimized}) were proposed to guide compiler auto-tuning through evolutionary. In addition, Perez et al.~\cite{perez2017automatic} proposed Irace to generate new compiler sequences based on performance distribution. The unsupervised learning based techniques search and obtain a desired optimization sequence by actually compiling and executing the target program with a large number of generated optimization sequences, which is very time-consuming. 

\textbf{Supervised learning based techniques.} The supervised learning based techniques construct a prediction model by optimization sequences (with the corresponding runtime performance of the compiled program), and then explore the optimization space by some predefined heuristics. In particular, Agakov et al. ~\cite{agakov2006using} randomly selected a number of optimization sequences to build a prediction model, and generated new optimization sequences with Markovian. Later, Ashouri et al.~\cite{ashouri2018automatic} proposed a Bayesian network based technique, which first builds a prediction model, and then generates new sequences through a Bayesian process that gradually approximated the desired solution. \add{
Ameer et al.~\cite{DBLP:conf/cgo/Haj-AliAWSAS20} proposed to extract the loop structures of the program, embed these loops, and then dynamically determine the flag sequence for all the loops through deep reinforcement learning.
}
Recently, Chen et al.~\cite{chen2021efficient} proposed another Bayesian network based technique BOCA, which first builds a prediction model to distinguish impactful flags from less impactful flags, and then generates new sequences through a Bayesian process. 

Our proposed \methodName~is also a supervised learning based technique, but it is different from the existing supervised learning based techniques in two aspects. First, \methodName~builds a prediction model via multiple phases of learning so as to alleviate the existing cost issue. Second, \methodName~selects optimization sequences through an improved particle swarm optimization algorithm, which alleviates the performance issue by considering both global and local optimum. 

Moreover, the supervised learning based techniques usually use Gaussian models as prediction models. Due to the excessive number of compiler compilation flags, the models constructed by such techniques usually suffer from accuracy issues, which are addressed by the improved particle swarm optimization algorithm of our \methodName.

The most relevant work is \textbf{BOCA}~\cite{chen2021efficient}. In particular, BOCA first iteratively selects sequences to construct a prediction model, which is used to predict whether an optimization flag is impactful, then explores the search space for impactful flags and space for less impactful flags separately. \methodName~is also a supervised learning based technique, and it is similar to BOCA because both of them build a prediction model iteratively with a small number of data. However, the purpose of their prediction models are different. The model of BOCA is mainly to identify whether an optimization flag is impactful, whereas the model of \methodName~is to predict the runtime performance of an optimization sequence. Moreover, when building a prediction model, BOCA uses only the sequence with good runtime performance whereas our \methodName~uses sequences with diversity. 

\section{CONCLUSION}
\label{sec:conclude}
To alleviate human efforts on compiler tuning for a specific program, in the literature several compiler auto-tuning techniques have been proposed, but all suffer from cost and performance issues. In this paper, we propose \methodName, a multiple-phase learning based compiler auto-tuning technique. To alleviate the cost issue, \methodName~iteratively builds a prediction model with a very small number of data to predict the runtime performance of an optimization sequence instead of actual execution. To alleviate the performance issue, \methodName~designs an improved particle swarm optimization algorithm, which balances the local optimum and global optimum through modifying acceleration constants. According to the experimental results on GCC and LLVM, the proposed \methodName~achieves the best runtime performances in most target programs, significantly outperforming the state-of-art compiler auto-tuning techniques. Moreover, the ablation analysis also demonstrates the contribution of each component in \methodName.

\bibliographystyle{ACM-Reference-Format}
\bibliography{ref}

\appendix

\end{document}